\documentclass[aps,prd,twocolumn,floatfix,showpacs]{revtex4}
\usepackage{graphicx,amsfonts,amssymb,amsbsy}
\usepackage{pstricks}

\usepackage{graphicx}
\usepackage{slashed}
\usepackage{bm}
\usepackage{amssymb}

\begin{document}

\title{Critical spectrum of fluctuations for deconfinement at proto-neutron star cores}
\author{G. Lugones$^1$ and A. G. Grunfeld$^{2,3,4}$}
 \affiliation{ $^1$ Universidade Federal do ABC, Centro de Ci\^encias
 Naturais e Humanas, Rua Santa Ad\'elia, 166, 09210-170, Santo Andr\'e, Brazil\\
 $^2$ Department of Physics, Sultan Qaboos University, P.O.Box: 36 Al-Khode 123 Muscat, Sultanate of Oman \\
 $^3$ CONICET, Rivadavia 1917, (1033) Buenos Aires, Argentina.\\
 $^4$ Departamento de F\'\i sica, Comisi\'on Nacional de
 Energ\'{\i}a At\'omica, (1429) Buenos Aires, Argentina.}

\begin{abstract}
We study the deconfinement of hadronic matter into quark matter in
a protoneutron star focusing on the effects of the finite size on
the formation of just-deconfined color superconducting quark
droplets embedded in the hadronic environment. The hadronic phase
is modeled by the non-linear Walecka model at finite temperature
including the baryon octet and neutrino trapping. For quark matter
we use an $SU(3)_f$ Nambu-Jona-Lasinio model including color
superconductivity. The finite size effects on the just deconfined
droplets are considered in the frame of the multiple reflection
expansion. In addition, we consider that  just deconfined quark
matter is transitorily out of equilibrium respect to weak
interaction, and we impose color neutrality and flavor
conservation during the transition. We calculate self-consistently
the surface tension and curvature energy density of the quark
hadron inter-phase and find that it is larger than the values
typically assumed in the literature. The transition density is
calculated for drops of different sizes, and at different
temperatures and neutrino trapping conditions. Then, we show that
energy-density fluctuations are much more relevant for
deconfinement than temperature and neutrino density fluctuations.
We calculate the critical size spectrum of energy-density
fluctuations that allows deconfinement as well as the nucleation
rate of each critical bubble. We find that drops with any radii
smaller than 800 fm can be formed at a huge rate when matter
achieves the bulk transition limit of 5$-$6 times the nuclear
saturation density.
\end{abstract}


\maketitle
\section{Introduction}

The cores of massive neutron stars have densities that may favor
the nucleation of small droplets of quark matter, that under
appropriate conditions may grow converting a large part of the
star into quark matter (for recent work see
\cite{Lugones2010,Logoteta,Fraga,Hempel} and references therein).
Since the the equation of state above the nuclear saturation
density is still uncertain  the problem is usually analyzed within
a two-phase description in which a hadronic model valid around the
nuclear saturation density $\rho_{0}$ is extrapolated to larger
densities and a quark model that is expected to be valid only for
asymptotically large densities is extrapolated downwards. Within
this kind of analysis some work has been performed recently in
order to determine the effect of different hadronic and quark
equations of state, as well as the effect of temperature, neutrino
trapping and color superconductivity \cite{Lugones2010}.

An important characteristic of the deconfinement transition in
neutron stars, is that quark and lepton flavors must be conserved
during the process leaving just deconfined quark matter that is
transitorily out of equilibrium with respect to weak interactions
(see \cite{Lugones2010} and references therein). When color
superconductivity is included in the analysis together with flavor
conservation, it is found that the most likely configuration of
the just deconfined phase is two-flavor color superconductor (2SC)
provided the pairing gap is large enough \cite{Lugones2005}.

Since our previous studies were made in bulk, we shall now include
finite size effects in the description of the just deconfined
drops. To this end we employ the Nambu-Jona-Lasinio (NJL) model
for quark matter and include finite size effects within the
multiple reflection expansion (MRE) framework (see
\cite{Madsen-drop,Kiriyama1,Kiriyama2} and references therein).
Through this analysis we can determine the density of hadronic
matter at which deconfinement is possible for different radii of
the just formed quark drops using typical conditions expected in
the interior of protoneutron stars (PNSs), i.e. temperatures in
the range of $0-60$ MeV and chemical potentials of the trapped
neutrino gas up to $200$ MeV. Notice that to form a finite size
drop it is needed some over-density with respect to the bulk
transition density in order to compensate the surface and
curvature energy cost. Since this energy cost depends on the drop
radius $R$, so does the necessary over-density necessary to
nucleate it. Thus, we can derive a critical fluctuation spectrum
($\delta \rho/ \rho$ versus $R$) delimiting which over-densities
can deconfine and grow unlimitedly and which ones will shrink back
to hadronic matter.

The paper is organized as follows. In Section II we briefly
describe main features of the hadronic model. In Sect. III we
present the quark model we use taking into account finite size
effects and derive self-consistently the surface tension and the
curvature energy from the MRE thermodynamic potential. In Sect IV
we the find the density at which color superconducting droplets of
different radii can be nucleated in protoneutron star matter. In
Section V we find the critical spectrum of density fluctuations as
well as the nucleation rate of the critical size droplets.  In
Sect. V we present our conclusions.

\section{The hadronic phase}
\label{hphase}

The hadronic phase is modeled by the non-linear Walecka
model (NLWM) \cite{qhd,GM1,boguta,Menezes} including the whole
baryon octet. The Lagrangian of the model is given by
\begin{equation}
{\cal L}={\cal L}_{B}+{\cal L}_{M}+{\cal L}_{L}, \label{octetlag}
\end{equation}
where the indices $B$, $M$ and $L$ refer to baryons, mesons and
leptons respectively. For the baryons we have
\begin{eqnarray}
{\cal L}_B= \sum_B \bar \psi_B \bigg[\gamma^\mu\left
(i\partial_\mu - g_{\omega B} \ \omega_\mu- g_{\rho B} \ \vec \tau
\cdot \vec \rho_\mu \right)
\nonumber\\
-(m_B-g_{\sigma B} \ \sigma)\bigg]\psi_B,
\end{eqnarray}
with $B$ extending over the nucleons $N= n$, $p$ and the following
hyperons $H = \Lambda$, $\Sigma^{+}$, $\Sigma^{0}$, $\Sigma^{-}$,
$\Xi^{-}$, and $\Xi^{0}$. The contribution of the mesons $\sigma$,
$\omega$ and $\rho$ is given by
\begin{eqnarray}
{\cal L}_{M} &=& \frac{1}{2} (\partial_{\mu} \sigma \ \!
\partial^{\mu}\sigma -m_\sigma^2 \ \! \sigma^2) - \frac{b}{3} \ \!
m_N\ \! (g_\sigma\sigma)^3 -\frac{c}{4} \ (g_\sigma \sigma)^4
\nonumber\\
& & -\frac{1}{4}\ \omega_{\mu\nu}\ \omega^{\mu\nu} +\frac{1}{2}\
m_\omega^2 \ \omega_{\mu}\ \omega^{\mu}
\nonumber \\ & &
-\frac{1}{4}\ \vec \rho_{\mu\nu} \cdot \vec \rho\ \! ^{\mu\nu}+
\frac{1}{2}\ m_\rho^2\  \vec \rho_\mu \cdot \vec \rho\ \! ^\mu,
\end{eqnarray}
where the coupling constants are
\begin{eqnarray}
g_{\sigma B}=x_{\sigma B}~ g_\sigma,~~g_{\omega B}=x_{\omega B}~
g_{\omega},~~g_{\rho B}=x_{\rho B}~ g_{\rho}
\end{eqnarray}
with $x_{\sigma B}$, $x_{\omega B}$ and $x_{\rho B}$ equal to $1$
for the nucleons and  0.9 for hyperons. Electrons and neutrinos are
included as a free Fermi gas in chemical equilibrium with all other particles.

There are five constants in the model that are determined by the
properties of nuclear matter, three that determine the nucleon
couplings to the scalar, vector and vector-isovector mesons
$g_{\sigma}/m_{\sigma}$, $g_{\omega}/m_{\omega}$,
$g_{\rho}/m_{\rho}$, and two that determine the scalar self
interactions $b$ and $c$. It is assumed that all hyperons in the
octet have the same coupling than the $\Lambda$. These couplings
are expressed as a ratio to the nucleon couplings mentioned above,
that we thus simply denote $x_\sigma$, $x_\omega$ and $x_\rho$.

In the present work we use the following values for the constants:
$(g_{\sigma}/m_{\sigma})^2 = 11.79 \, \textrm{fm}^2$ ,
$(g_{\omega}/m_{\omega})^2 = 7.149 \, \textrm{fm}^2$,
$(g_{\rho}/m_{\rho})^2 =  4.411 \, \textrm{fm}^2$,    $b =
0.002947 $ and  $c = -0.001070$. This model has been labelled as
GM4 in previous work \cite{Lugones2010}.  The equation of state is 
rather stiff and gives a maximum mass of 2 M$_\odot$, that seems to be more
adequate in light of the recently determined mass of the pulsar PSR J1614-2230 
with $ M =  1.97 \pm 0.04 M_\odot$ \cite{Demorest}. This
is the largest mass reported ever for a pulsar with a high
precision. For more details on the explicit form of the equation of state the reader
is referred to Ref. \cite{Lugones2010}.

\section{The quark matter phase}
\label{qphase}

The just deconfined quark matter phase is implemented by using the
$SU(3)_f$ NJL effective model with the inclusion quark-quark
interactions, which are the responsible for color
superconductivity.

The corresponding Lagrangian is given by
\begin{eqnarray}
{\cal L} &=& \bar \psi \left(i \rlap/\partial - \hat m \right) \psi \nonumber \\
& & + G \sum_{a=0}^8 \left[ \left( \bar \psi \ \tau_a \ \psi \right)^2 + \left( \bar \psi \ i \gamma_5 \tau_a \ \psi \right)^2 \right]
\label{lagrangian} \\
& & + 2H \!\! \sum_{A,A'=2,5,7} \left[ \left( \bar \psi \ i \gamma_5 \tau_A \lambda_{A'} \ \psi_C \right) \left( \bar \psi_C \ i \gamma_5 \tau_A \lambda_{A'} \ \psi \right) \right] \nonumber
\end{eqnarray}
where $\hat m=\mathrm{diag}(m_u,m_d,m_s)$ is the current mass
matrix in flavor space. In the present paper we work in the
isospin symmetric limit $m_u=m_d=m$. Moreover, $\tau_i$ and
$\lambda_i$ with $i=1,..,8$ are the Gell-Mann matrices
corresponding to the flavor and color groups respectively, and
$\tau_0 = \sqrt{2/3}\ 1_f$. Finally, the charge conjugate spinors
are defined as follows: $\psi_C = C \ \bar \psi^T$ and $\bar
\psi_C = \psi^T C$, where $\bar \psi = \psi^\dagger \gamma^0$ is
the Dirac conjugate spinor and $C=i\gamma^2 \gamma^0$.

In order to determine the relevant thermodynamic quantities we
have to obtain the grand canonical thermodynamic potential at
finite temperature $T$ and chemical potentials $\mu_{fc}$. Here,
$f=(u,d,s)$ and $c=(r,g,b)$ denotes flavor and color indices
respectively. In the following we present the thermodynamic
potential for the bulk system and then, in Sect. III B, we discuss
the effects of finite size in the effective potential, for
the spherical droplets.

\subsection{Quark matter in bulk}

Starting from Eq. (\ref{lagrangian}), we perform the usual
bosonization of the theory. We introduce the scalar and
pseudoscalar meson fields $\sigma_a$ and $\pi_a$ respectively,
together with the bosonic diquark field $\Delta_A$. In what
follows we will work within the mean field approximation (MFA), in
which these bosonic fields are expanded around their vacuum
expectation values and the corresponding fluctuations neglected.
For the meson fields this implies $\hat \sigma = \sigma_a \tau_a =
\textrm{diag}(\sigma_u,\sigma_d,\sigma_s)$ and $\pi_a=0$.
Concerning the diquark mean field, we will assume that in the
density region of interest only the 2SC phase might be relevant.
Thus, we adopt the ansatz $\Delta_5 = \Delta_7 = 0$, $\Delta_2 =
\Delta$. Integrating out the quark fields and working in the
framework of the Matsubara and Nambu-Gorkov formalism we obtain
the following MFA quark thermodynamic potential
$\Omega^{MFA}_q(T,\mu_{fc},\sigma_u,\sigma_d,\sigma_s,|\Delta|)$ per unit volume
(further calculation details can be found in Refs.
\cite{Huang:2002zd,Ruester:2005jc,Blaschke:2005uj,ciminale})
\begin{eqnarray}
\frac{\Omega_q^{MFA}}{V} =  2 \int_0^\Lambda \frac{k^2dk }{2  \; \pi^2} \sum_{i=1}^9  \omega(x_i,y_i) +  \nonumber\\
\frac{1}{4G}(\sigma_u^2+\sigma_d^2+\sigma_s^2)  +
\frac{|\Delta|^2}{2H},
\end{eqnarray}
where $\Lambda$ is the cut-off of the model and $\omega(x,y)$ is defined by
\begin{eqnarray}
\omega(x,y) = - [ x + T\ln[1+e^{-(x-y)/T}]  \nonumber\\
+  T\ln[1+e^{-(x+y)/T}]  ] ,
\end{eqnarray}
with
\begin{eqnarray}
x_{1,2} & = & E , \quad  x_{3,4,5} = E_s  ,   \nonumber\\
x_{6,7} & = & \sqrt{\bigg[ E +  \frac{(\mu_{ur} \pm \mu_{dg})}{2} \bigg]^2 + \Delta^2}  ,  \nonumber\\
x_{8,9} & = & \sqrt{\bigg[ E +  \frac{(\mu_{ug}  \pm \mu_{dr})}{2} \bigg]^2 + \Delta^2}  ,  \nonumber \\
y_{1}    & = & \mu_{ub} , \quad  y_2 = \mu_{db} , \quad  y_{3} = \mu_{sr}  , \nonumber \\
y_{4}    & = & \mu_{sg}  , \quad  y_{5} = \mu_{sb} ,      \nonumber\\
y_{6,7} & = & \frac{(\mu_{ur}-\mu_{dg})}{2}  ,  \quad y_{8,9} = \frac{\mu_{ug} - \mu_{dr}}{2} .
\end{eqnarray}
Here, $E = \sqrt{ k^2+ M^2}$ and $E_s=\sqrt{  k ^2 + M_s^2}$,
where $M_f = m_f + \sigma_f$. Note that in the isospin limit we
are working $\sigma_u = \sigma_d = \sigma$ and, thus, $M_u = M_d =
M$.

The total thermodynamic potential  of the quark matter phase (super-index Q) is obtained by adding to $\Omega_q^{MFA}$ the
contribution of the leptons and a vacuum constant. Namely,
\begin{equation}
\Omega^{Q} = \Omega_q^{MFA} + \Omega_e + \Omega_{\nu_e} - \Omega_\textrm{\tiny vac} \label{QMP}
\end{equation}
where $\Omega_e$ and $\Omega_{\nu_e}$ are the thermodynamic
potentials of the electrons and neutrinos respectively. For them
we use the expression corresponding to a free gas of
ultra-relativistic fermions
\begin{equation}
\frac{\Omega_l(T,\mu_{l}) }{V} = - P_l =  - \gamma_l \left( \frac{\mu_l^4}{24 \pi^2} +
\frac{\mu_l^2T^2}{12} + \frac{7\pi^2T^4}{360} \right), \nonumber
\end{equation}
where $P$ stands for the pressure, $l = e, \nu_e$, and the
degeneracy factor is $\gamma_e = 2$ for electrons and
$\gamma_{\nu_e} = 1$ for neutrinos. Notice that in Eq. (\ref{QMP})
we have subtracted the constant $\Omega_\textrm{\tiny vac} \equiv
- P_\textrm{\tiny vac} V$ in order to have a vanishing pressure at
vanishing temperature and chemical potentials. More details
are shown below in Sec. \ref{parametrizations}.

\subsection{Finite size effects: inclusion of MRE}

In the present work we consider the formation of finite size
droplets of quark matter. The effect of finite size is included in
the thermodynamic potential adopting the formalism of multiple
reflection expansion (MRE; see Refs. \cite{Madsen-drop,Kiriyama1,Kiriyama2} and references therein).

In the MRE framework, the  modified  density of states of a finite spherical droplet is given by
\begin{equation}
\rho_{MRE}(k,m_f,R) = 1 + \frac{6\pi^2}{kR}
f_S\left(\frac{k}{m_f}\right) + \frac{12\pi^2}{(kR)^2}
f_C\left(\frac{k}{m_f}\right)
 \end{equation}
where
\begin{equation}
f_S \left(\frac{k}{m_f} \right) = - \frac{1}{8 \pi} \left(1 -
\frac{2}{\pi} \arctan \frac{k}{m_f} \right)
\end{equation}
and
\begin{equation}
f_C \left(\frac{k}{m_f} \right) =  \frac{1}{12 \pi^2} \left[1 -
\frac{3k}{2m_f} \left(\frac{\pi}{2} - \arctan \frac{k}{m_f}
\right)\right]
\end{equation}
are the surface and curvature contributions to the new density of
states respectively. For $f_C$ we employ the Madsen ansatz
\cite{Madsen-drop} because its functional form for any finite
quark mass has not yet been derived in the MRE frame.

As shown in \cite{Kiriyama2}, the density of states of MRE for
massive quarks is reduced  compared with the bulk one, and for a
range of small momentum becomes negative. The way of excluding
this non physical negative density of states is to introduce an
infrared cutoff in momentum space (see \cite{Kiriyama2} for
details). Thus, in our quark model, which includes the MRE
formalism for finite spherical droplets of color superconducting
quark matter, we have to perform the following replacement
\begin{equation}
\int_0^{\Lambda} \frac{k^2 \, dk}{2 \pi^2}  \rightarrow
\int_{\Lambda_{IR}}^\Lambda \frac{k^2 \, dk}{2 \pi^2} \rho_{MRE}.
\label{MRE}
\end{equation}
In order to obtain the value of $\Lambda_{IR}$, we have to solve
the equation $\rho_{MRE} = 0$ with respect to the momentum $k$ and
take the larger root as the IR cut-off. Note that $\rho_{MRE}$
depends on the quark mass and the radius of the droplets and
consequently the $\Lambda_{IR}$ has the same dependence (more
details are given below in Sec. \ref{parametrizations}).

Therefore, the full thermodynamic potential reads:
\begin{eqnarray}
\frac{\Omega^Q_{{MRE}}}{V} & = & 2
\int_{{{\Lambda_{IR}}}}^\Lambda \frac{k^2dk }{2 \; \pi^2}
\rho_{MRE} \sum_{i=1}^9  \omega(x_i,y_i)    \nonumber \\
   & &  + \frac{1}{4G}(\sigma_u^2+\sigma_d^2+\sigma_s^2)  + \frac{|\Delta|^2}{2H}  \nonumber \\
   & &  - P_e  - P_{\nu_e} + P_\textrm{\tiny vac} .
\label{fullomega}
\end{eqnarray}
Multiplying on both sides of the last equation by the volume of
the quark matter drop and rearranging terms we arrive to the
following form for $\Omega^Q_{MRE}$
\begin{equation}
\Omega^Q_{{MRE}} = -P^Q V + \alpha S + \gamma C ,
\end{equation}
where the pressure $P^Q$ is given by
\begin{eqnarray}
P^Q   \equiv  - \frac{\partial \Omega^Q_{{MRE}}}{ \partial V }
\bigg|_{T, \mu, S, C}  & = & - 2 \int_{{{\Lambda_{IR}}}}^\Lambda
\frac{k^2dk }{2  \; \pi^2} \sum_{i=1}^9 \omega(x_i,y_i)  \nonumber\\
 &  &- \frac{1}{4G}(\sigma_u^2+\sigma_d^2+\sigma_s^2)  - \frac{|\Delta|^2}{2H}  \nonumber\\
 & &  + P_e  + P_{\nu_e} - P_\textrm{\tiny vac} ,
\end{eqnarray}
the surface tension is
\begin{equation}
\alpha \equiv  \frac{\partial \Omega^Q_{{MRE}}}{ \partial S }
\bigg|_{T, \mu, V, C}  = 2 \int_{{{\Lambda_{IR}}}}^\Lambda k
dk  f_S \sum_{i=1}^9 \omega(x_i,y_i) ,
\label{surfacetension}
\end{equation}
and the curvature energy density is
\begin{equation}
\gamma \equiv  \frac{\partial \Omega^Q_{{MRE}}}{ \partial C }
\bigg|_{T, \mu, V, S}  = 2 \int_{{{\Lambda_{IR}}}}^\Lambda  dk
f_C \sum_{i=1}^9 \omega(x_i,y_i).
\label{curvatureenergy}
\end{equation}
Here we are considering a spherical drop, i.e.  the area is $S= 4\pi R^2$  and the curvature is
$C=8\pi R$.

From the grand thermodynamic potential  $\Omega^{Q}_{MRE}$ we can
readily obtain the number density of quarks of each flavor and
color $n_{fc} \equiv  - V^{-1}{ \partial
\Omega^{Q}_{{MRE}}}/{\partial \mu_{fc} }$, the number density of
electrons  $n_{e} = - V^{-1} {\partial \Omega^{Q}_{{MRE}}
}/{\partial \mu_e}$, and the number density of electron neutrinos
$n_{\nu_e} = - V^{-1} {\partial \Omega^{Q}_{{MRE}} }/{\partial
\mu_{\nu_e}}$. The corresponding number densities of each flavor,
$n_f$, and of each color, $n_c$, in the quark phase are given by
$n_f = \sum_{c} n_{fc}$  and  $n_c = \sum_{f} n_{fc}$
respectively.  The baryon number density reads $n_B = \frac{1}{3}
\sum_{fc} n_{fc} = (n_u + n_d + n_s)/3$. Finally, the Gibbs free
energy per baryon is
\begin{equation}
g = \frac{1}{n_B}\left(\sum_{fc}  \mu_{fc} \ n_{fc} + \mu_e \ n_e +  \mu_{\nu_e} \ n_{\nu_e} \right).
\label{g_quark}
\end{equation}

\begin{table}[t!]
\centering
\begin{tabular}{c|ccccc}
\hline\hline & $m_{u,d} $ [MeV] & $m_s$ [MeV] & $\Lambda$ [MeV] & $G\Lambda^2$& $H/G$\\
\hline
Set 1 & 5.5 & 112.0  & 602.3 & 4.638 & 3/4 \\
Set 2 & 5.5 & 110.05 & 631.4 & 4.370 & 3/4 \\
\hline
\end{tabular}
\caption{The two sets of NJL parameters.} \label{sets}
\end{table}

\subsection{Parametrizations}
\label{parametrizations}

For the NJL model, the values of the quark masses and the coupling
constant $G$ can be obtained from the meson properties in the
vacuum. Here we use a set of parameters taken from
\cite{Hatsuda:1994pi}, but without the 't Hooft flavor mixing
interaction. The procedure, obtained from \cite{Buballa2005}, is
to keep $\Lambda$ and $m$ fixed and then tune the remaining
parameters $G$ and $m_s$ in order to reproduce $M =367.6$ MeV and
$M_s=549.5$ MeV at zero temperature and density. An estimate of
$H/G$ can be obtained from Fierz transformation of the
one-gluon-exchange interactions in which case one gets $H/G =$
0.75. The resulting parameter sets are given in Table \ref{sets}.
For this set of parameters we get $\Omega_\textrm{\tiny vac} / V =
-P_\textrm{\tiny vac} = -4301$ MeV/fm$^3$ and $-5099$ MeV/fm$^3$
(for set 1 and 2 respectively).

As we previously mentioned, the value of $\Lambda_{IR}$ is the
largest root when solving $\rho_{MRE} = 0$ with respect to $k$,
depending on $m_f$ and R. For a given mass, we find numerically
that $\Lambda_{IR}(R)$ can be fitted as follows
\begin{equation}
\Lambda_{IR} = a \, R^b \label{LIR}
\end{equation}
with $R$ in fm and $\Lambda_{IR}$ in MeV. The coefficients $a$ and
$b$ are found to be $135.45$ and $-0.85$ for $m_u = 5.5$ MeV;
227.72 and $-0.86$ for $m_s = 110.5$ MeV; 228.60 and $-0.87$ for
$m_s = 112$ MeV.

\subsection{Flavor conservation, color neutrality and other conditions}

In order to derive a quark matter EOS from the above formulae it
is necessary to impose a suitable number of conditions on the
variables  $\{\mu_{fc}\}, \mu_e, \mu_{\nu_e},\sigma, \sigma_s$ and
$\Delta$. Three of these conditions are consequences from the fact
that the thermodynamically consistent solutions correspond to the
stationary points of $\Omega^Q_{{MRE}}$ with respect to
$\sigma$, $\sigma_s$, and $\Delta$. Thus, we have
\begin{eqnarray}
\frac{\partial\Omega^{Q}_{{MRE}}}{\partial\sigma} =0, \qquad
\frac{\partial\Omega^{Q}_{{MRE}}}{\partial\sigma_s} =0, \qquad
\frac{\partial\Omega^{Q}_{{MRE}}}{\partial|\Delta|}=0.
\label{gapeq}
\end{eqnarray}
To obtain the remaining conditions one must specify the physical
situation in which one is interested in. As in previous works
\cite{Madsen-drop,deconf1,deconf2,Lugones2010}, we are dealing
here with just deconfined quark matter that is temporarily out of
chemical equilibrium under weak interactions. The appropriate
condition in this case is flavor conservation between hadronic and
deconfined quark matter. This can be written as
\begin{equation}
Y^H_f = Y^Q_f   \;\;\;\;\;\; f=u,d,s,e, \nu_e \label{flavor}
\end{equation}
being $Y^H_f \equiv n^H_f / n^H_B$ and  $Y^Q_i \equiv n^Q_f /
n^Q_B$ the abundances of each particle in the hadron and quark
phase respectively. It means that the just deconfined quark phase
must have the same ``flavor'' composition than the $\beta$-stable
hadronic phase from which it has been originated. Notice that,
since the hadronic phase is assumed to be electrically neutral,
flavor conservation ensures automatically the charge neutrality of
the just deconfined quark phase.  The conditions given in Eq.
(\ref{flavor}) can be combined to obtain
\begin{eqnarray}
n_d & =  & \xi ~ n_u,  \label{h3}\\
n_s  & =  & \eta ~ n_u , \\
n_{\nu_e}  & =  &  \kappa ~ n_u ,  \\
3 n_{e} &  =  & 2 n_{u} - n_{d} - n_{s} ,
\end{eqnarray}
\noindent where $n_i$ is the particle number density of the
$i$-species in the quark phase. The quantities $\xi \equiv Y^H_d /
Y^H_u$, $\eta \equiv Y^H_s / Y^H_u$  and $\kappa \equiv
Y^H_{\nu_e} / Y^H_u$ are functions of the pressure and
temperature, and they characterize the composition of the hadronic
phase. These expressions are valid for \textit{any} hadronic EOS.
For hadronic matter containing $n$, $p$, $\Lambda$, $\Sigma^{+}$,
$\Sigma^{0}$, $\Sigma^{-}$, $\Xi^{-}$ and $\Xi^{0}$, we have
\begin{eqnarray}
\xi &=& \frac{n_p  +  2  n_n  + n_{\Lambda} + n_{\Sigma^{0}} +  2
n_{\Sigma^{-}}  + n_{\Xi^{-}}}{2  n_p  +  n_n  +  n_{\Lambda} + 2
n_{\Sigma^{+}} + n_{\Sigma^{0}}  +  n_{\Xi^{0}}}, \label{xi} \\
\eta &=& \frac{n_{\Lambda}  + n_{\Sigma^{+}} + n_{\Sigma^{0}}  +
n_{\Sigma^{-}} + 2 n_{\Xi^{0}} + 2 n_{\Xi^{-}}}{2  n_p  +  n_n  +
n_{\Lambda} + 2 n_{\Sigma^{+}} + n_{\Sigma^{0}}  +  n_{\Xi^{0}}}, \\
\kappa &=& \frac{n^H_{\nu_e}}{2  n_p  +  n_n  + n_{\Lambda} + 2
n_{\Sigma^{+}} + n_{\Sigma^{0}}  +  n_{\Xi^{0}}}. \label{eta}
\end{eqnarray}
Additionally, the deconfined phase must be locally colorless; thus
it must be composed by an equal number of red, green and blue
quarks
\begin{equation}
n_r = n_g = n_b
 \label{colorless}.
\end{equation}
Also, $ur$, $ug$, $dr$, and $dg$ pairing will happen provided that
$|\Delta|$ is nonzero, leading to
\begin{equation}
n_{ur}=n_{dg}  , \quad n_{ug}=n_{dr}. \label{pairing}
\end{equation}

In order to have all Fermi levels at the same value, we consider
\cite{Lugones2005}
\begin{eqnarray}
n_{ug} = n_{ur} , \quad n_{sb} = n_{sr}.
\label{equalfermilevels}
\end{eqnarray}
These two equations, together with Eqs. (\ref{colorless}) and
(\ref{pairing}) imply that $n_{ur}=n_{ug}=n_{dr}=n_{dg}$ and
$n_{sr}=n_{sg}=n_{sb}$ \cite{Lugones2005}.

Finally, including the conditions Eqs.(\ref{gapeq}) we have 13
equations involving the 14 unknowns ($\sigma$, $\sigma_s$,
$|\Delta|$, $\mu_e$, $\mu_{\nu_e}$ and $\{\mu_{fc}\}$). For given
value of one of the chemical potentials (e.g. $\mu_{ur}$), the set
of equations can be solved once the values of the parameters
$\xi$, $\eta$, $\kappa$, the temperature $T$ {{and the radius of
the drop $R$}} are given. Instead of $\mu_{ur}$, we can provide a
value of the Gibbs free energy per baryon $g_\textrm{\scriptsize
quark}$ and solve simultaneously Eqs.
(\ref{h3})-(\ref{equalfermilevels}) together with Eqs.
(\ref{gapeq}) in order to obtain $\sigma$, $\sigma_s$, $|\Delta|$,
$\mu_e$, $\mu_{\nu_e}$ and $\{\mu_{fc}\}$.

\section{Deconfinement of color superconducting droplets in proto-neutron star matter}

The total thermodynamic potential of a quark matter drop immersed
in an homogeneous hadronic environment is $\Omega = \Omega^H  +
\Omega^Q_{{MRE}} $, where the indexes $H$ and $Q$ refer to the
hadronic and the quark phase respectively. For the hadronic phase
we have $\Omega^H = - P^H V^H $ and for the quark phase we have
{{$\Omega^Q_{{MRE}}= -P^Q V^Q + \alpha S + \gamma C$}}.

The condition of mechanical equilibrium is given by
\cite{Landau,Madsen-drop}:
\begin{equation}
\frac{\partial \Omega}{\partial R} \bigg|_{T,\mu, V} =
\frac{\partial \Omega^H}{\partial R} \bigg|_{T,\mu, V} +
\frac{\partial \Omega^Q_{{MRE}}}{\partial R} \bigg|_{T,\mu, V}
= 0.
\end{equation}
Since $V = V^Q + V^H$ is constant,  we have:
\begin{equation}
 -P^Q \frac{dV^Q}{dR} + \alpha \frac{dS}{dR} + \gamma \frac{dC}{dR} + P^H \frac{dV^Q}{dR} =  0.
\end{equation}

Thus, for a spherical droplet the condition for mechanical equilibrium reads (c.f. \cite{Madsen-drop}):
\begin{equation}
P^Q  - \frac{2 \alpha}{R}   -   \frac{2 \gamma}{R^2}  - P^H  = 0.
\label{mech_equil}
\end{equation}
Notice that in the bulk limit $R \rightarrow \infty$ we find the standard condition $P^H = P^Q$.

Additionally, we assume thermal and chemical equilibrium, i.e. the
Gibbs free energy per baryon are the same for both hadronic matter
and quark matter at a given common temperature. Thus, we have
\begin{eqnarray}
g^H = g^Q \; , \qquad T^H = T^Q \; .
\label{gibbs}
\end{eqnarray}
If we fix the radius $R$ of the deconfined drop for a given
temperature $T^H$ and neutrino chemical potential of the trapped
neutrinos in the hadronic phase $\mu_{\nu_e}^H$, there is an
unique hadronic pressure $P^H$ at which the equilibrium conditions
are fulfilled. Notice that, differently than in the bulk case
studied in \cite{Lugones2010}, $P^H$ and $P^Q$ are not equal
because of the appearance of a surface and a curvature term in the
condition for mechanical equilibrium. Also, the mass-energy
density $\rho^H$ and $\rho^Q$ at the equilibrium point are
different in general. Similarly, while the abundance $Y_{\nu_e}$
of neutrinos is the same in both the hadronic and just deconfined
quark phases, the chemical potentials $\mu_{\nu_e}^Q$ and
$\mu_{\nu_e}^H$ are different.

\begin{figure*}[t]
\includegraphics[scale=0.45]{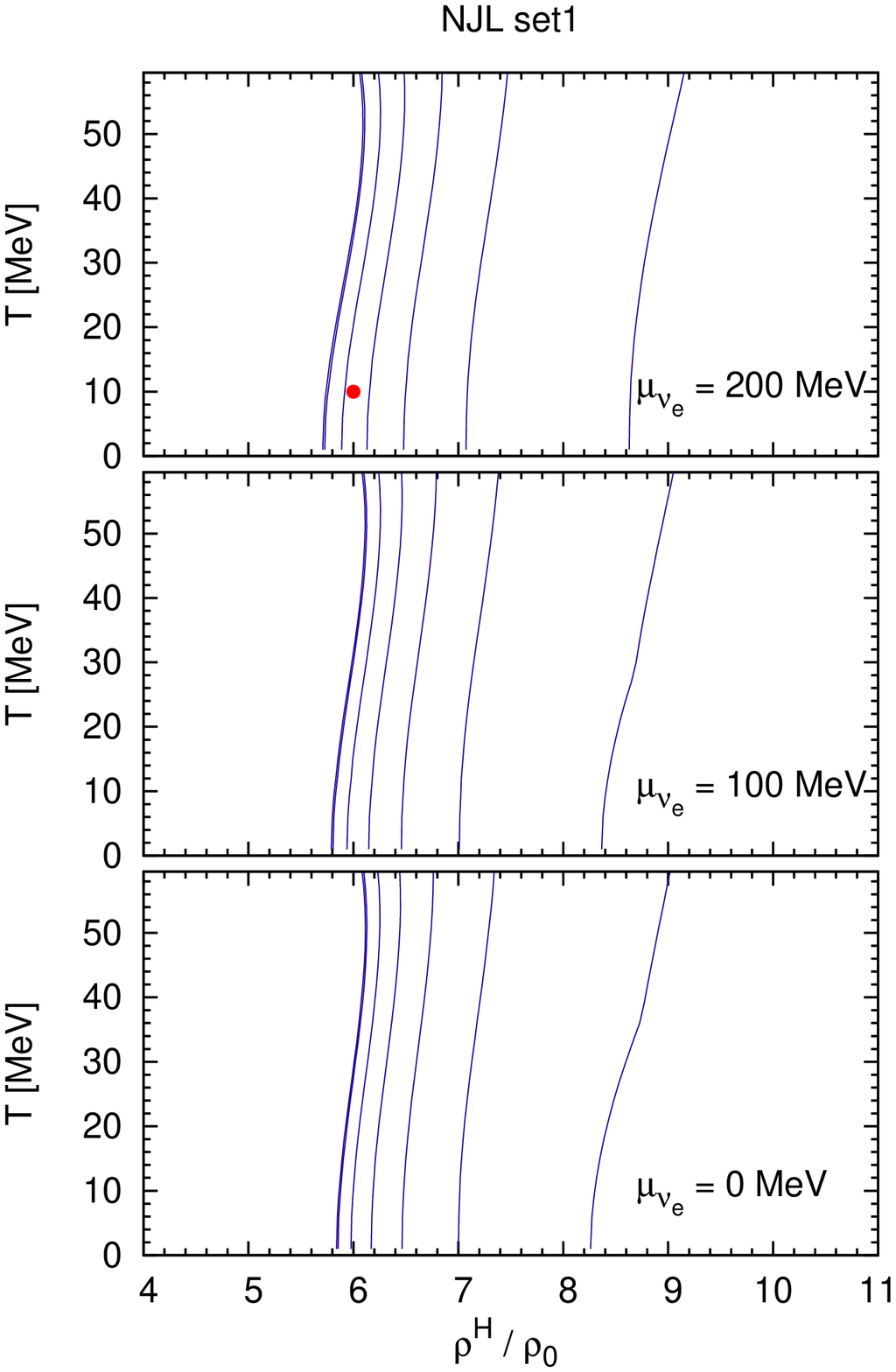}
\includegraphics[scale=0.45]{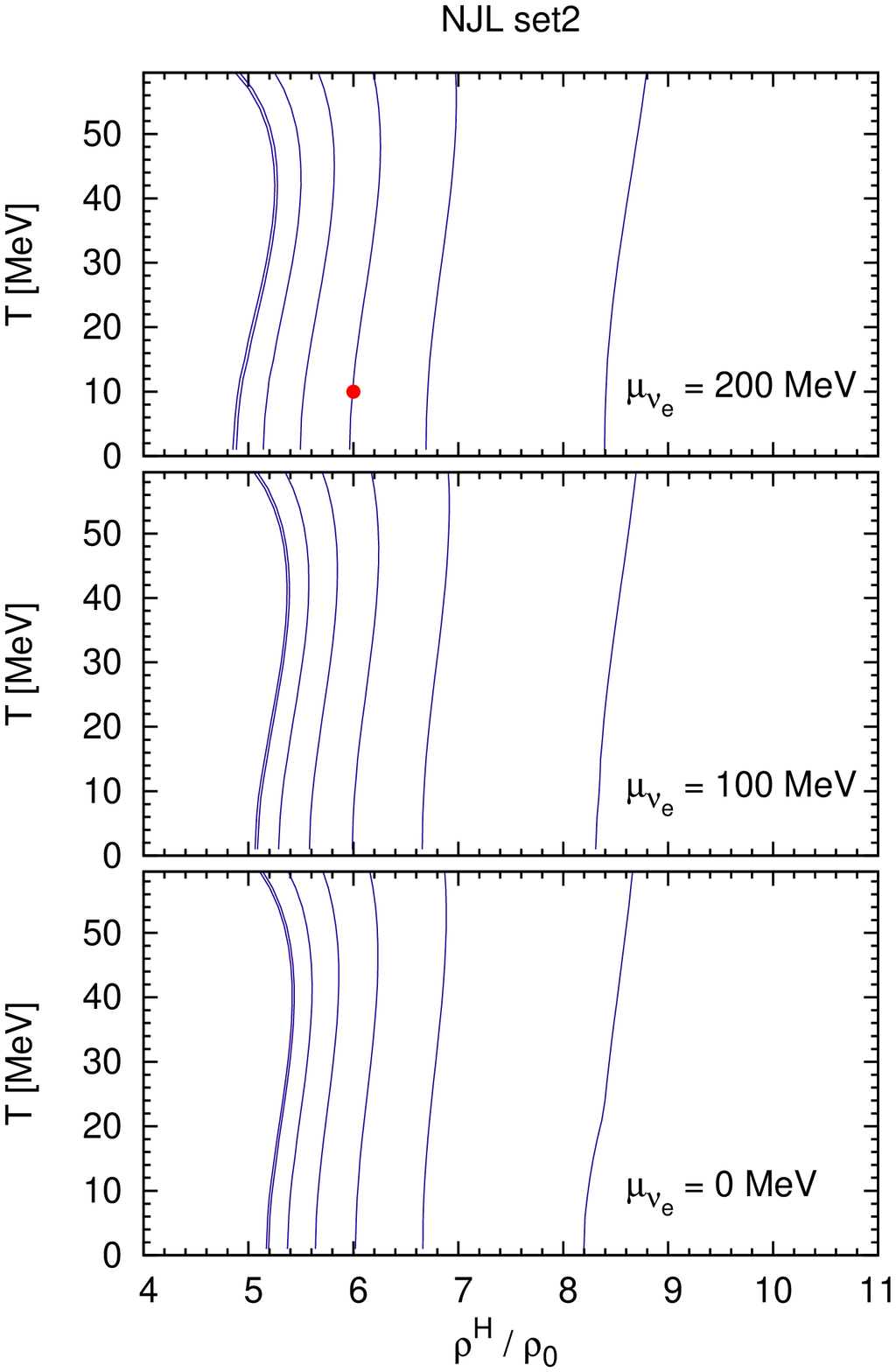}
\caption{For different temperatures we show the mass-energy
density $\rho^H$ of hadronic matter  above which it is
energetically favorable the conversion of the hadronic phase into
deconfined drops of radius $R$.  $\rho_0 =  2.7 \times 10^{14}$ g
cm$^3$ is the nuclear saturation density. For the hadronic phase
we employ the model introduced in Sec. II. For the quark phase we
use set 1 of the NJL model presented in Sec III in the left
panels, and set 2 in the right panels. In each figure the neutrino
chemical potential in the hadronic phase is fixed to the following
values: $\mu_{\nu_e}^H =0, 100, 200$ MeV. Within each plot, each
curve correspond to a different value of $R$. From right to left
we have $R [\textrm{fm}]= 2, 5, 10, 20, 50, 500, \infty$. Notice
that in all figures the curves for the bulk case ($R= \infty$) are
almost coincident with the curves for $R = 500$ fm. A possible
interpretation of these figures is best illustrated with an
example: let us consider a hadronic system with a density $\rho^H
\approx 6 \times \rho_0$ at a temperature $T = 10$  MeV and with
trapped neutrinos having a chemical potential $\mu_{\nu_e}^H =
200$ MeV. This state is represented by a circle in the two upper
panels of this figure. Looking at the position of the circle with
respect to the curves corresponding to different radii, we see
that it is energetically favorable to convert hadronic matter into
quark drops provided they have a radii $R \gtrsim 50$ fm (if quark
matter is described by the NJL set 1 EoS) and $R \gtrsim 10$ fm
(using NJL set 2 EoS). That is, radii whose curves lie to the left
of the circle are energetically favored while those on the right
are forbidden because of the energy cost of forming a surface. A
more realistic interpretation of the figure is given in Sec. V.}
\label{fig1}
\end{figure*}

\begin{figure*}[t]
\begin{center}
\includegraphics[scale=0.45]{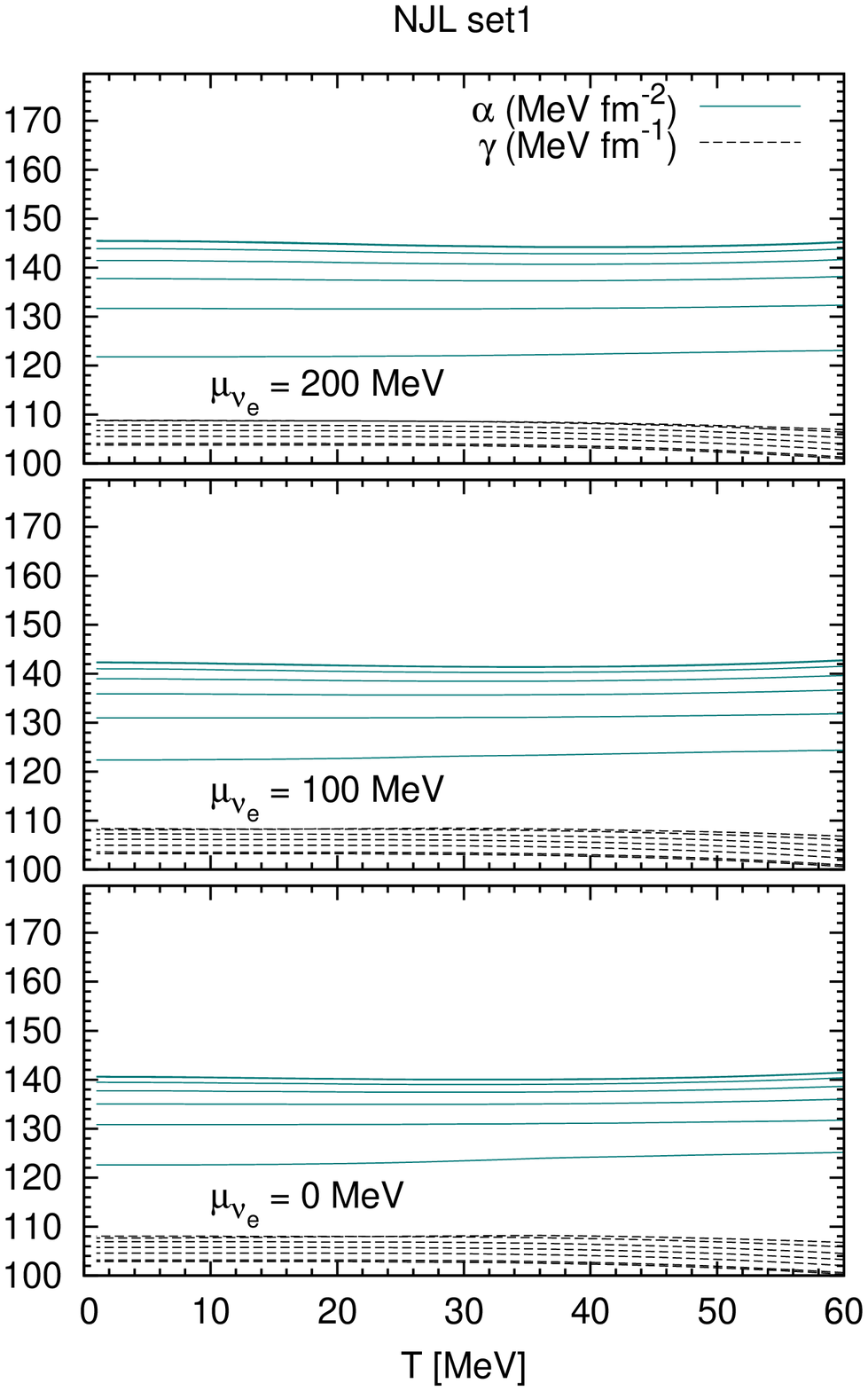}
\includegraphics[scale=0.45]{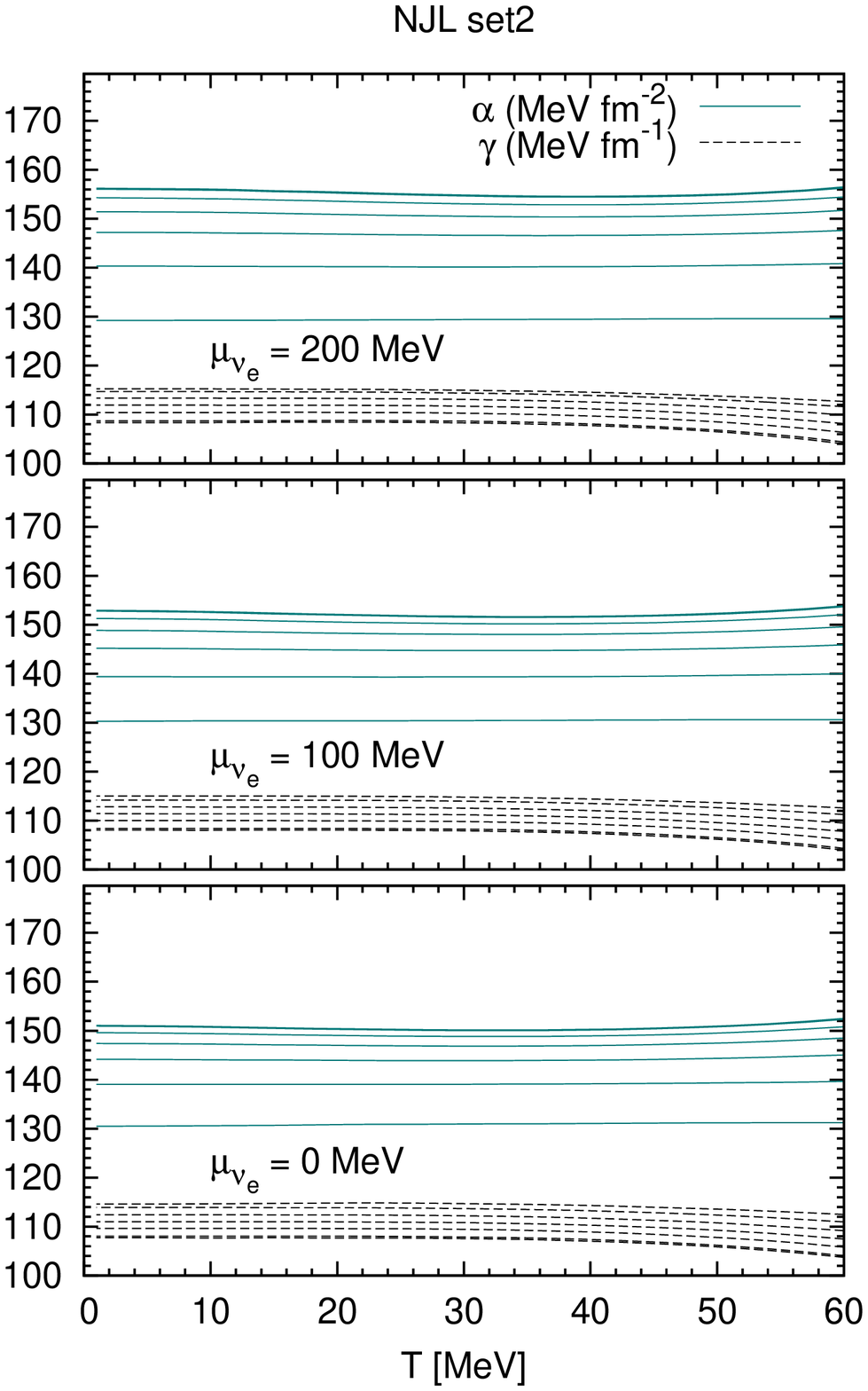}
\caption{ The surface tension $\alpha$ and the curvature energy
$\gamma$ as a function of temperature for different radii of the
droplet and with the set of $\{ \mu_{fc},\sigma, \sigma_f, \Delta
\}$ that arises from the equilibrium conditions leading to Fig. 1.
Different curves correspond to different radii $R$; in the case of
$\alpha$ we have $R [\textrm{fm}]= 2, 5, 10, 20, 50, 500, \infty$
from bottom to top while in the case of $\gamma$ the values are
the same but from top to bottom.   Both $\alpha$ and $\gamma$ are
roughly independent of the temperature but there is a larger
dependence with $R$.} \label{alpha}
\end{center}
\end{figure*}

In the present work we want to describe thermodynamic conditions analogous to those encountered
in protoneutron stars. From numerical simulations
of the first tens of seconds of evolution \cite{PNS},
we know that the protoneutron star cools from $T
\sim 40$ MeV to temperatures below 2-4 MeV in about one minute.
In that period, the chemical potential $\mu_{\nu_e}^H$ of the trapped neutrinos
decreases from $\sim 200$ MeV to essentially zero.
Then,  taking into account typical protoneutron star conditions, we have solved
Eqs. (\ref{mech_equil}) and (\ref{gibbs}) together with the
conditions showed in Sections II and III, for temperatures in the
range $0-60$ MeV, $\mu_{\nu_e}^H$ in the range $0-200$ MeV and for
different radii of the color superconducting droplets.  The results are
displayed in Figs. 1 and 2 for the parameterizations of the
equations of state given in previous sections.

In Fig. 1  we show the mass-energy density of hadronic matter (in
units of the nuclear saturation density $\rho_0 =  2.7 \times
10^{14}$ g cm$^3$) above which it is energetically favorable the
conversion of a portion of hadronic matter into an identical
deconfined drop (i.e. with the same $T$, $P$, $g$, $Y_f$ and
radius $R$). For the calculations we considered the following
possibilities: set 1 of the NJL model in the left panels and set 2
in the right panels, temperatures in the range 0-60 MeV, and
neutrino chemical potential  $\mu_{\nu_e}^H =0, 100, 200$ MeV.
Within each plot we consider different values of $R$; from right
to left the curves correspond to $R [\textrm{fm}]= 2, 5, 10, 20,
50, 500, \infty$. Notice that, in the six panels of Fig. 1, the
curves for the bulk case ($R= \infty$) are almost coincident with
the curves for $R = 500$ fm.  The main observed effect in Fig. 1
is that the transition density increases considerably for small
radii. This is a consequence of the significant  increase of the
surface tension  $\alpha$ and the curvature energy $\gamma$ for
small $R$ (see Fig. 2).

In Fig. 2 we display the surface tension $\alpha$ and the
curvature energy $\gamma$ as a function of temperature for
different radii of the droplet and with the set of  $\{
\mu_{fc},\sigma, \sigma_f, \Delta \}$ that arises from the
equilibrium condition discussed before. Notice that $\alpha$ and
$\gamma$ do not depend explicitly on the radius $R$ of the drop
(see Eqs. (\ref{surfacetension}) and (\ref{curvatureenergy})), but
for a given $R$, $T$ and $\mu^H_{\nu_e}$ there is a unique  set of
$\{ \mu_{fc}, \sigma, \sigma_f, \Delta \}$ satisfying the
equilibrium conditions, i.e. $\alpha$ and $\gamma$ depend on $R$
through  $\{ \mu_{fc},\sigma, \sigma_f, \Delta \}$.  From Fig. 2
we see that while both $\alpha$ and $\gamma$ are roughly
independent of the temperature, there is a larger dependence with
the radius $R$  leading to the behavior of the transition density
explained in the caption of Fig. 1. Notice also that the
here-found values of $\alpha$ are larger that those found within the MIT Bag model
which are typically $\sim$ 30-60 MeV fm$^{-2}$ for drops larger than few fm \cite{Wen2010}.

\section{Fluctuations and deconfinement}

\begin{figure*}
\begin{center}
\includegraphics[scale=0.44]{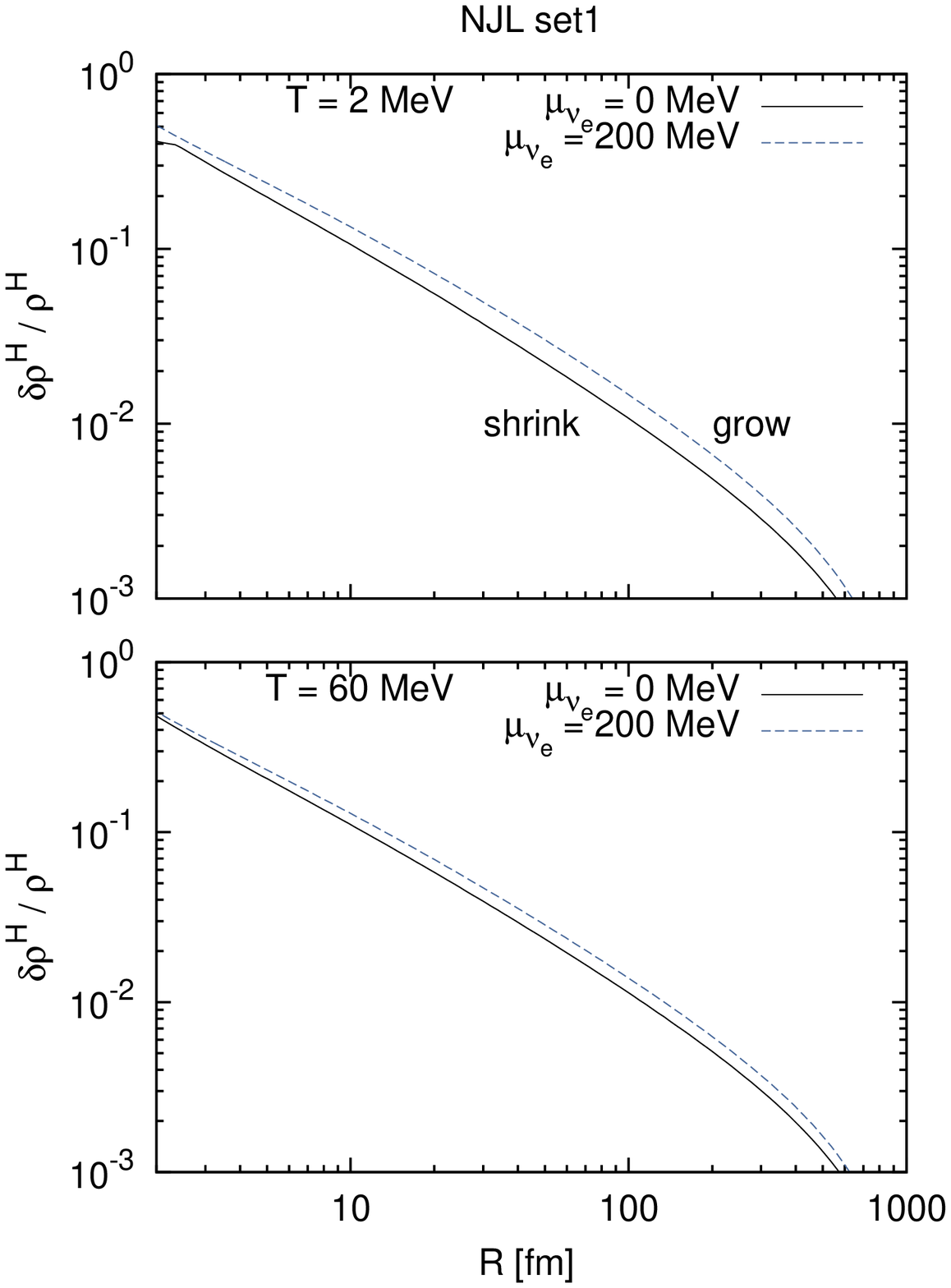}
\includegraphics[scale=0.44]{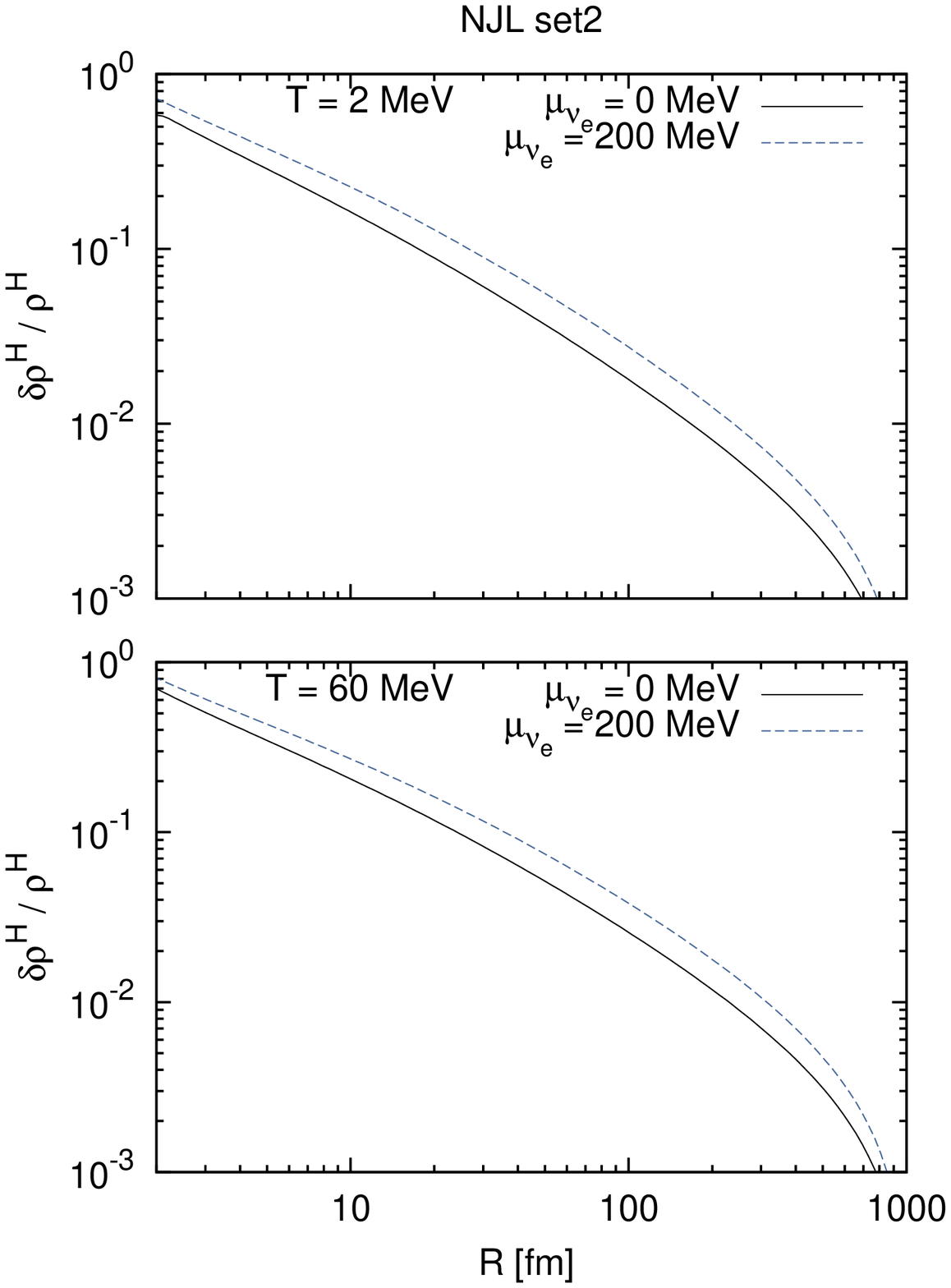}
\caption{Critical spectrum for fluctuations that allow
deconfinement of hadronic matter in a protoneutron star. As in the previous figures,  the results
were calculated using GM4 + NJL \textit{set1} on the left panel and  GM4 + NJL \textit{set 2} on the right panel.
The spectrum does not depend significantly on the temperature and the chemical potential
of trapped neutrinos. Fluctuations in hadronic matter having a given $\delta \rho^H / \rho^H$ are able to grow
if they have a size $R$ larger than  the here-shown critical one.  }
\end{center}
\end{figure*}

\begin{figure*}
\begin{center}
\includegraphics[scale=0.44]{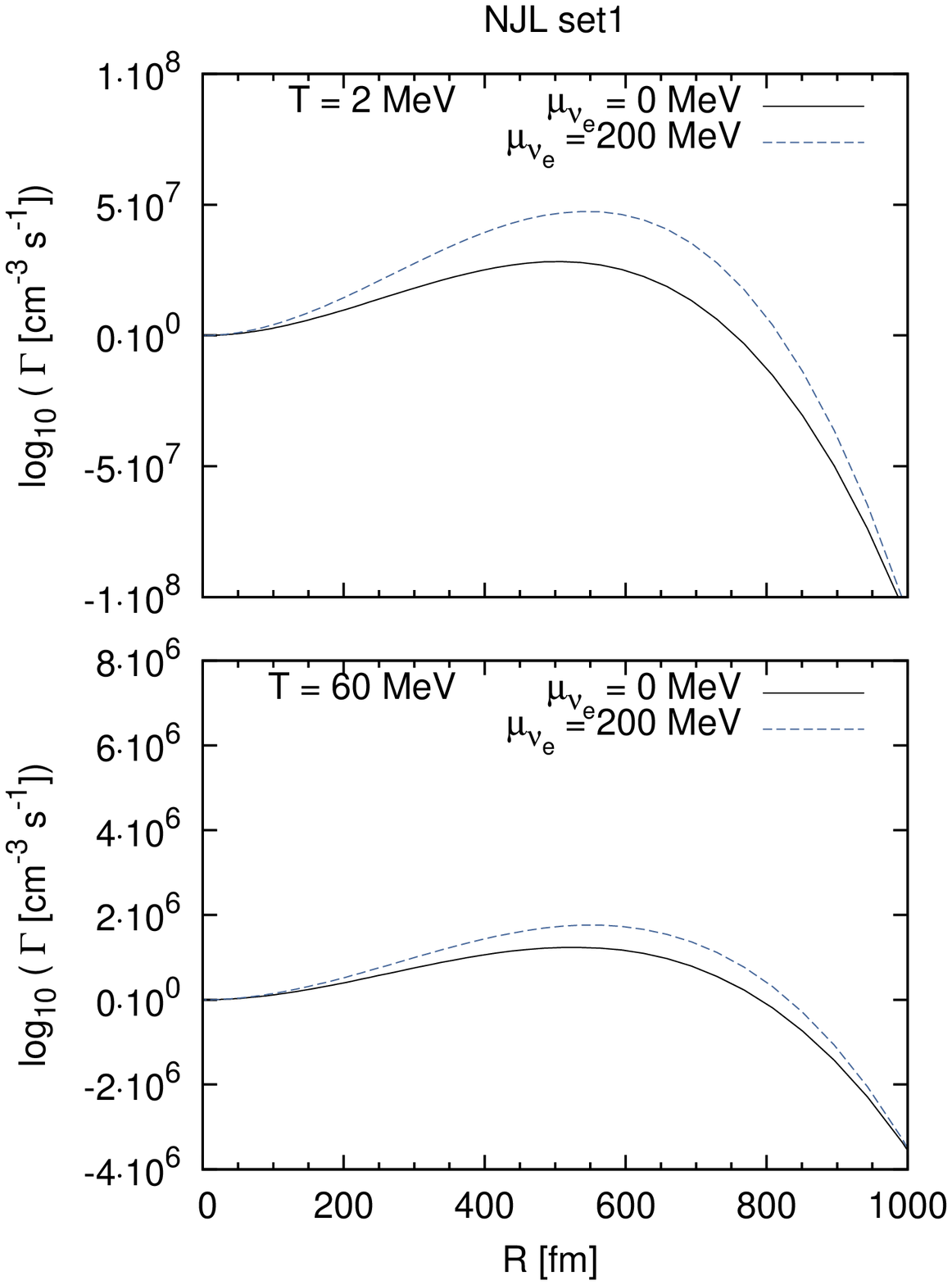}
\includegraphics[scale=0.44]{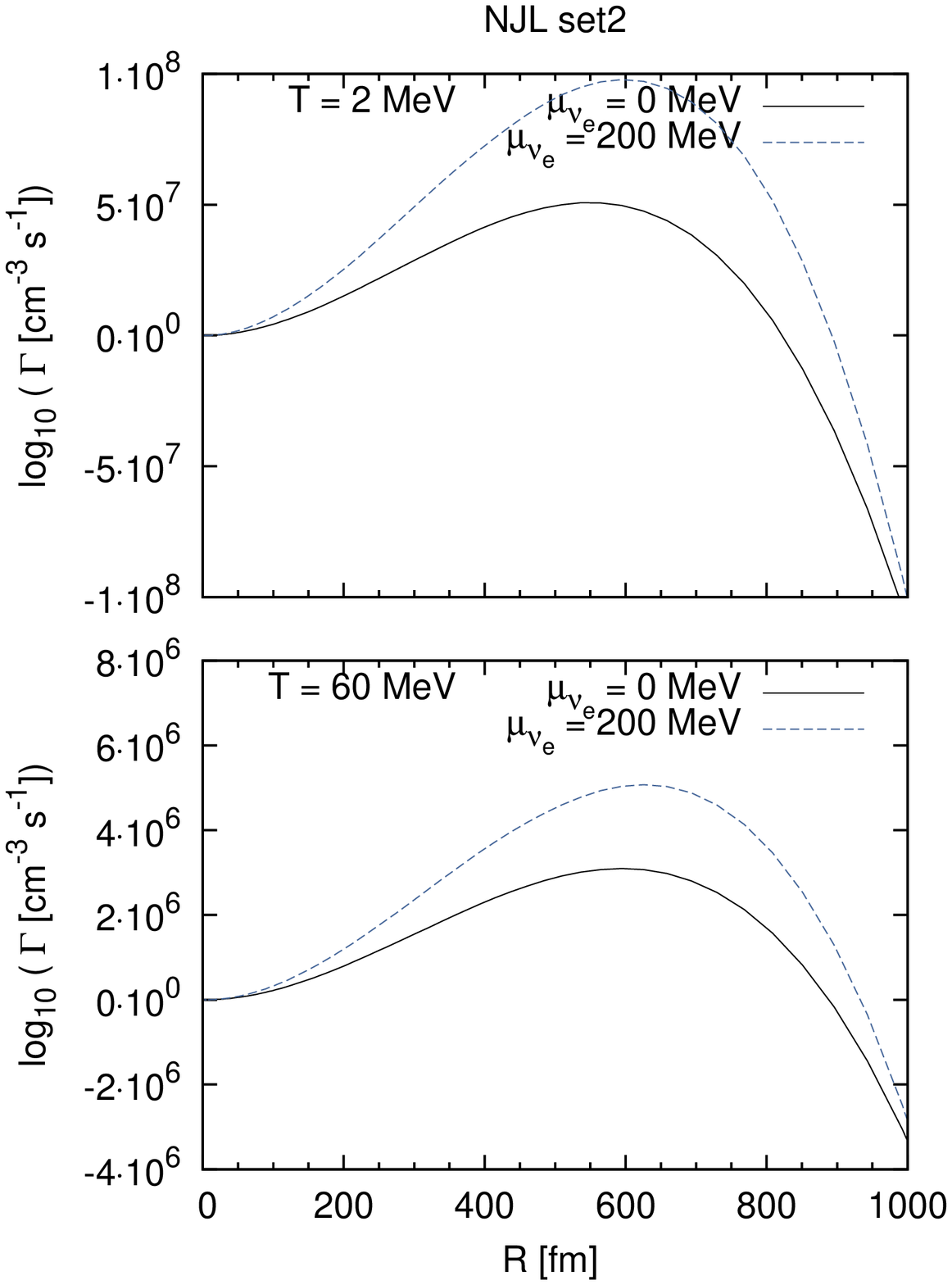}
\caption{Nucleation rate for bubbles of the critical size.   }
\end{center}
\end{figure*}

According to the theory of homogeneous nucleation, the free energy
involved in the formation of a spherical quark bubble of radius
$R$ is given by \cite{Madsen-nucl}
\begin{equation}
\Delta \Omega =   - {{4 \pi} \over 3} R^3 \Delta P + 4 \pi \alpha R^2  + 8 \pi \gamma R ,
\label{work}
\end{equation}
where $\Delta P = P^Q - P^H$ is the pressure difference between
internal and external parts of the bubble. For given $\Delta P$,
$\alpha$ and $\gamma$,  the extremal points (maximum or minimum)
of $\Delta \Omega$ are obtained from $\partial  \Delta \Omega /
\partial R = 0$, which leads to Eq. (\ref{mech_equil}). Thus, the
critical radii are given by:
\begin{equation}
R_{\pm} = {\alpha \over{\Delta P}}\left( 1 \pm \sqrt{1 + b}\right) ,
\label{rcrit}
\end{equation}
with $b \equiv 2\gamma \Delta P/ \alpha^2$.

For $b < -1$ both solutions are complex and $\Delta \Omega$ is a
monotonically decreasing function of $R$. This means that any
small fluctuation of one phase into the other will gain energy by
expanding and a rapid phase transition is likely to occur.  For $b
\geq -1$,  $\Delta \Omega$  has a local minimum at $R_-$ and a
local maximum at $R_+$.  For $\Delta \Omega (R_+) <  0 $ we have
again that any small fluctuation is energetically favored. For
$\Delta \Omega (R_+) >0$ bubbles with radii larger than $R_+$ gain
energy by growing unlimitedly, while those below the critical size
gain energy by shrinking to zero (if $R_- <0$) or to $R_-$ (if
$R_- >0$ ). In this case, the standard assumption in the theory of
bubble nucleation in first order phase transitions is that bubbles
form with a critical radius $R_+$ \cite{Madsen-nucl}.

The approach we adopted in the previous section is closely related
to what we explained in the above paragraph. Instead of finding
the critical radius for arbitrary values of $\Delta P$, $\alpha$
and $\gamma$, we fixed $R$ and found the corresponding $\Delta P$,
$\alpha$ and $\gamma$ that satisfy the conditions presented in
Secs. II, III and IV. Since Eq. (\ref{mech_equil}) is satisfied by
construction, the radius $R$ is precisely the critical radius
$R_+$ introduced in Eq. (\ref{rcrit}), given that we choose the
solution that verifies $\partial^2  \Delta \Omega / \partial R^2 <
0$.

In the light of the previous discussion we may give another
interpretation to the results presented in Fig. 1. Let us
consider, for simplicity, a uniform hadronic system with trapped
neutrinos having a chemical potential $\mu_{\nu_e}^H$ and a
constant mass-energy density  \textit{infinitesimally to the
right} of the curve with $R = \infty$ in the $T -  \rho^H/\rho_0$
plane corresponding to the same  $\mu_{\nu_e}^H$. Since this
density corresponds to the bulk transition we shall refer to it as
$\rho_{bulk}^H(T,\mu_{\nu_e}^H)$.   For such a density, it is
favorable to convert the system into quark matter \textit{in
bulk}, but this is not possible in practice due to the surface and
curvature energy cost.

However, fluctuations in the independent thermodynamic
variables $\{T, \rho^H,\mu_{\nu_e}^H \}$ of the hadronic fluid may
drive some portion of it  to a state described by $\{T + \delta T,
\rho^H + \delta \rho^H,\mu_{\nu_e}^H + \delta \mu_{\nu_e}^H  \}$.
Nevertheless, notice that the curves of Fig. 1 are quite vertical,
i.e. the transition is not very sensitive to changes in $T$. The
same holds for variations in $\mu_{\nu_e}^H$. Thus, we shall
consider only energy-density fluctuations with radius $R_{fl}$
that drive some part of the hadron fluid to a density
$\rho_{*}^{H}  = \rho_{bulk}^H + \delta \rho^H$ to the right of a
curve with a given $R$. As explained in the caption of Fig. 1,
quark drops with radii larger than $R$ are energetically favored.
Thus, if the fluctuation has a size $R_{fl}$ larger than $R$ it
will be energetically favorable for it to grow indefinitely.  In
order to quantify this, we calculate the difference $\delta
\rho^H$ between $\rho_{bulk}^H$ and the hadronic density of the
point that allows nucleation with radius $R$ or larger. In such a
way we can construct  a critical spectrum  $\delta \rho^H/\rho^H$
as a function of $R$ for different values of $T$ and
$\mu_{\nu_e}^H$ as seen in Fig. 3. Fluctuations of a given
over-density $\delta \rho^H/\rho^H$ must have a size larger than
the critical value given in Fig. 3 in order to grow. Equivalently,
fluctuations of a given size must have an over-density $\delta
\rho^H/\rho^H$ larger than the critical one for that size.

We can calculate the formation rate of critical bubbles through
\begin{equation}
\Gamma \approx T^4 \exp (-\delta \Omega_c/T).
\label{prob}
\end{equation}
where in our case $\delta \Omega_c$ is the work required to form a
quark bubble with the critical radius from hadronic matter at the
bulk transition point
\begin{equation}
\delta \Omega_c \equiv   - {{4 \pi} \over 3} R^3 (P^Q - P^H_{bulk} ) + 4 \pi \alpha R^2  + 8 \pi \gamma R ,
\label{wcrit}
\end{equation}
Instead of $T^4$, different prefactors are used for
$\Gamma$ in other works (see e.g. \cite{Logoteta}). However, this
fact does not affect significantly the results because
$\Gamma$ is largely dominated by the exponent in Eq. (\ref{prob});
i.e. we always have $\log_{10} \Gamma \approx \log_{10}
(\mathrm{prefactor})  -  \delta \Omega_c/[ T  \ln(10)]$ with the
second term much larger than the first.

The results are given in Fig. 4 and show that critical bubbles
with $R\gtrsim 800$ fm are strongly disfavored while those with $R
\lesssim 800$ fm have a huge rate. In practical situations, i.e.
at neutron star cores, this means that if fluctuations lead
hadronic matter to the bulk transition point,  quark drops with $R
\lesssim 800$ fm  will nucleate instantaneously.

\section{Summary and Conclusions}

In the present paper we have studied the deconfinement of quark
matter in protoneutron stars employing a  two-phase description of
the first order phase transition. For the hadronic phase we used
the non-linear Walecka model (Sect. II) which includes the whole
baryon octet, electrons and trapped electron neutrinos in
equilibrium under weak interactions. For the just deconfined quark
matter we used a $SU(3)_f$ NJL model including color
superconducting quark-quark interactions. In this phase, finite
size effects are included via the multiple reflection expansion
framework (Sect. III). We considered that color superconducting
quark droplets are formed in mechanical and thermal equilibrium
with the hadronic environment when the Gibbs free energy of both
phases are equal. Also, since deconfinement is a strong
interaction process, we consider that the just formed quark phase
has the same flavor composition than the hadronic $\beta$-stable
phase, and consequently it is out of chemical equilibrium under
weak interactions. The further $\beta$-equilibration of such quark
phase is not addressed within the present paper.

Through this model we determined the density of hadronic matter at
which deconfinement is possible for different radii of the just
formed quark drops, as well as for different values of the
temperature and the chemical potential of the trapped neutrinos
(see Fig. 1). We have used typical conditions expected in the
interior of protoneutron stars, i.e. temperatures in the range of
$0-60$ MeV and chemical potentials of the trapped neutrino gas up
to $200$ MeV. As expected, we find that the transition density
increases significantly when the radius of the droplet decreases.
This is a consequence of the increase of the surface tension
$\alpha$ and the curvature energy  $\gamma$ for small radii.
Additionally, we found that $\alpha$ and $\gamma$ at the
deconfinement point are almost independent of the temperature and
the neutrino chemical potential (see Fig. 2).

When the bulk transition density is reached in the core of a
neutron star, it is energetically favored to convert a
macroscopically large portion of hadronic matter into quark
matter. But in practice, it is needed some over-density with
respect to the bulk transition density in order to compensate the
surface and curvature energy cost of a finite drop.  Since this
energy cost depends on the drop radius, so does the necessary
over-density and over-pressure necessary to nucleate it. Thus, we
can derive a critical fluctuation spectrum $\delta \rho^H/\rho^H$
versus $R$ delimiting which fluctuations are able to grow
unlimitedly and which will shrink (see Fig. 3). Typically,
fluctuations of $\delta \rho^H/\rho^H \sim 0.001-0.1$ above the
bulk point are needed for the nucleation of drops with $R \sim
10-1000$ fm. However, the nucleation rates $\Gamma$ vary over
several orders of magnitude. Our results show that drops with $R
\sim 2-800$ fm have a huge nucleation rate while those with $R
\gtrsim 800$ fm are strongly suppressed (see Fig. 4).

We have also shown that fluctuations in the temperature and in the
chemical potential of trapped neutrinos are not very important for
deconfinement. This is in contrast with previous results found
within the frame of the MIT Bag model (e.g. in
\cite{Horvath} it is argued that nucleation is suppressed
at $T \lesssim 2$ MeV and in \cite{deconf1} it is found that neutrino
trapping precludes deconfinement). Instead, fluctuations in the
energy density are the more efficient way to trigger the
transition.

Notice that the nucleation rate and the typical radii of
deconfined drops are also very different from the values found
within the MIT Bag model (see e.g. \cite{Logoteta} and references
therein). The drops studied in \cite{Logoteta} have
typically radii less than 10 fm and a long nucleation time, in
contrast ours may have much larger radii and nucleate almost
instantaneously. This is due to the use of different equations of
state for the quark phase as well as for the different treatments
of the surface and curvature terms. While in \cite{Logoteta} the
surface tension is assumed to be constant ($\alpha$ = 30 MeV
fm$^{-2}$), in our work $\alpha$ and $\gamma$ are calculated self
consistently within the MRE formalism resulting non-constant
values around 140 MeV fm$^{-2}$ and 110 MeV fm$^{-1}$
respectively. Since our surface tension is larger, larger critical
drops are obtained. The typical $|\delta \Omega_c|$ is also larger
and results in huge nucleation rates. In the context of
protoneutron stars the main conclusion is that if the bulk
transition point is attained near the star centre, quark matter
drops with $R \lesssim 800$ fm will nucleate instantaneously.
Since the bulk transition density is $\sim 5-6 \rho_0$, this
should happen for stars with masses larger than $\sim 1.5 - 1.6
M_{\odot}$.

\end{document}